\documentclass[aps,twocolumn,showpacs]{revtex4}

\usepackage{graphicx}
\usepackage{amsmath,amssymb}
\usepackage{amsfonts}
\usepackage{color}
\usepackage{ulem}

\begin{document}

\title{Experimental inhibition of decoherence on flying qubits via bang-bang control}

\author{S.~Damodarakurup}
\affiliation{Dipartimento di Fisica, Universit\`a di Camerino, via Madonna delle Carceri, 9, I-62032 Camerino (MC), Italia.}

\author{M.~Lucamarini}
\affiliation{Dipartimento di Fisica, Universit\`a di Camerino, via Madonna delle Carceri, 9, I-62032 Camerino (MC), Italia.}

\author{G.~Di~Giuseppe}
\affiliation{Dipartimento di Fisica, Universit\`a di Camerino, via Madonna delle Carceri, 9, I-62032 Camerino (MC), Italia.}

\author{D.~Vitali}
\affiliation{Dipartimento di Fisica, Universit\`a di Camerino, via Madonna delle Carceri, 9, I-62032 Camerino (MC), Italia.}

\author{P.~Tombesi}
\affiliation{Dipartimento di Fisica, Universit\`a di Camerino, via Madonna delle Carceri, 9, I-62032 Camerino (MC), Italia.}

\begin{abstract}
Decoherence may significantly affect the polarization state of
optical pulses propagating in dispersive media because of
the unavoidable presence of more than a single frequency in the
envelope of the pulse. Here we report on the suppression of polarization decoherence in a ring cavity obtained by properly retooling
for photonic qubits the ``bang-bang'' protection technique
already employed for nuclear spins and nuclear-quadrupole qubits.
Our results show that
bang-bang control can be profitably extended to
quantum information processes involving flying polarization
qubits.
\end{abstract}


\date{\today}

\maketitle

%%%---------------------------------------------------------%%%
%%%-------------------- Introduction --------------------%%%
%%%---------------------------------------------------------%%%

The struggle against environmental decoherence has a long history,
which proceeds from the refocusing techniques of Nuclear Magnetic
Resonance (NMR) spectroscopy~\cite{Ernst1987} to Quantum
Error-Correcting Codes
(QECC)~\cite{Shor1995,Steane1999,Chiaverini2004,Boulant2005},
Decoherence-Free Subspaces
(DFS)~\cite{Zanardi1997,Kielpinski2001,Prevedel2007}, Quantum
Feedback (QF)~\cite{Vitali1997}, and dynamical ``Bang-Bang'' (BB)
decoupling~\cite{Viola1998,Viola99,Vitali1999,Zanardi1999,Viola2001,Kofman2004,Fraval2005,Facchi2005}.
Despite the impressive achievements of these techniques, QECC and
DFS require a large amount of extra resources~\cite{Steane1999},
while QF is limited by measurement inefficiencies. In dynamical BB
decoupling, the system undergoes a sequence of suitably tailored
unitary operations which do not require ancillas or measurements.
The physical idea behind BB comes from refocusing techniques of
NMR spectroscopy~\cite{Ernst1987}: control cycles are implemented
in time via a sequence of strong and rapid pulses that provide a
full decoupling from the environment (and all its undesired
effects) if the controls are applied faster than the bath
correlation time. The decoherence suppression results in the
increase of the NMR ``transversal'' relaxation time $T_2$, which
is related to dephasing~\cite{Morton2006}. Besides NMR, dynamical
decoupling has been suggested for inhibiting the decay of unstable
atomic states~\cite{Agarwal2001a,Kofman2004}, suppressing the
decoherence of magnetic states~\cite{Search2000a}, and reducing
the heating in ion traps~\cite{Vitali2002}.
\\\indent
In this Letter we apply the BB technique to a flying qubit,
specifically the polarization state of a
pulse circling in a ring cavity. In fact, the properties of
optical elements (here the mirror reflectivity) depend both on
frequency and polarization and this, in conjunction with a finite
integration time of the detectors, results in a trace over the
frequency degree of freedom and effectively spoils the coherence
of a polarization state. A dephasing process takes place in the
cavity, and after few round-trips the polarization state is almost
completely mixed. On the contrary, an arbitrary polarization state
is preserved for many round-trips when the BB controls, realized
by suitably oriented wave-plates, are inserted in the optical
path.\\
\indent Our experiment provides a proof-of-principle demonstration
that BB control can be profitably extended to quantum
communication protocols involving flying polarization qubits. In
fact, a generic communication channel can be divided into many
small portions with constant dispersive properties. For example,
in a single-mode optical fiber, birefringence is nearly constant
over lengths much smaller than the fiber
beat-length~\cite{Galtarossa2000}. Few round-trips of our
ring-cavity well mimic one of such small portions, and we show
that BB decoupling efficiently inhibits any kind of polarization
decoherence in the ring-cavity. Therefore, even though dispersive
properties will vary randomly along the channel, provided that BB
controls are repeated on small enough distances, one expects that
the polarization qubit is going to be preserved for lengths much
longer than currently achieved~\cite{Massar2007}.
%
%%%---------------------------------------------------------%%%
%%%----------------------- Setup -------------------------%%%
%%%---------------------------------------------------------%%%
%
\vspace{-.2cm}
\begin{figure}[h!]
   \centering
   \includegraphics[width=.475\textwidth]{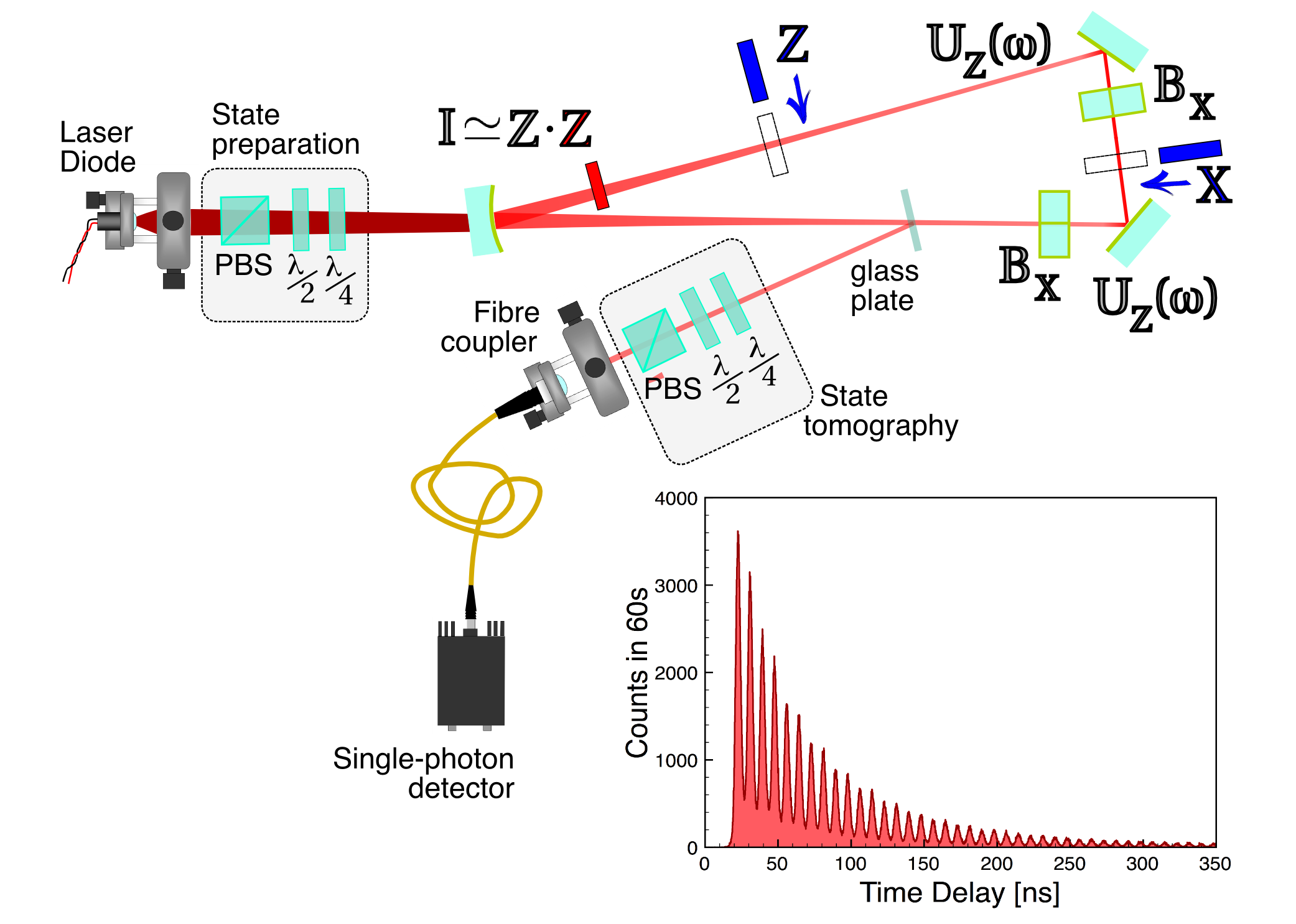}
   \caption{
   (Color online).
   Schematics of the experimental apparatus. A typical acquisition
   run, with a sequence of peaks, each corresponding to a cavity round-trip, is also shown.}
   \label{fig:BB_Setup}
\end{figure}
\\
\indent The experimental apparatus employed to demonstrate this
effect is depicted in Fig.~\ref{fig:BB_Setup}. A laser diode at
$\lambda_0 \simeq 800$~nm wavelength and bandwidth $\Delta \lambda
\simeq 15$~nm, is pulsed at repetition rate of 100~KHz and pulse
duration $\sim$~100~ps, and injected in a triangular ring cavity
through a spherical mirror with radius of curvature 1~m,
reflectivity 98\%. The state emerging from the laser is a
coherent state with average photon number per pulse $\mu$ about
equal to 1 soon after the mirror.
The cavity is also formed by two flat mirrors with reflectivity
higher than 99\%. The aperture angle of the cavity at the
spherical mirror is $8^\circ$, and the cavity length is $2.01$~m.
The polarization state of the pulses is prepared by using a polarizing beam-splitter (PBS) and a set of
$\lambda/2$ and $\lambda/4$ wave-plates. At every round trip the
light is extracted from the cavity with 4\% probability using a
100~$\mu$m thin glass plate.
%
%%%%%----------------APPROXIMATE SINGLE PHOTON PULSE---------------------%%%%%%%
Therefore the resulting coherent state entering the
detection apparatus has a mean photon number per pulse
$\mu\lesssim 4\times 10^{-2}$, and the probability of having two
or more photons in a detection event is less than 2\%, as we have
experimentally verified.
The polarization state is analyzed by means of the tomographic
technique~\cite{James2001}, and then sent into a multimode fibre
connected to a single-photon detector. The output signal from the
detector stops the time conversion in a Time-to-Amplitude
Converter (TAC) synchronized with the laser. The time delay
between the sync and output signals is then recorded by a
Multi-Channel Analyzer (MCA) with 8192 channels. The acquisition
electronics has a time-resolution of $102$~ps.  The typical
acquisition run is reported in the inset of
Fig.~\ref{fig:BB_Setup} and shows a sequence of peaks, each
corresponding to a cavity round-trip. The interval between
adjacent peaks amounts to 6.80~ns, in agreement with the
given cavity-length.\\

It is worth noticing that the experimental setup detailed
above justifies a description in terms of \textit{single-photon
polarization qubits}. The reason is twofold. On one side, although
the weak coherent state circling in the cavity may contain more
than one photon, each photon does not interact with the others, for all the
operations in the cavity involve only linear and passive optical
elements. On the other side, when the coherent state reaches the
detection apparatus, its intensity is so low that the probability
of a two-photon event is negligible. Hence the detection process
postselects, with high probability, only the single-photon pulses.
This allows us to write the state injected into the cavity, soon
after the spherical mirror, as an effective single-photon state
\begin{equation}\label{InEffState}
    |\psi \rangle_{in}^{eff} = \int d\omega\, \mathcal{E}(\omega)\,
|\omega \rangle\otimes |\pi\rangle_{in},
\end{equation}
where
$|\omega\rangle\otimes|\pi\rangle_{in}=\left[\alpha_H\hat{a}_H(\omega)^{\dagger}+\alpha_V
\hat{a}_V(\omega)^{\dagger}\right]|0\rangle $, with
$\hat{a}_S(\omega)^{\dagger}$ creating a photon with frequency
$\omega$ and linear polarization state
$S=\{|V\rangle,|H\rangle\}$, i.e. respectively orthogonal and
parallel to the plane of the cavity; $\mathcal{E}(\omega)$ is the
amplitude spectrum of the pulse, normalized such that $ \int
d\mu_\omega \equiv \int d\omega |\mathcal{E}(\omega)|^2 = 1$, and
$|\pi\rangle_{in}=\left(\alpha_H,\alpha_V\right)^{T}$ denotes the
frequency-independent input polarization state.

%%%---------------------------------------------------------%%%
%%%---------------------- ONLY CAVITY ----------------------%%%
%%%---------------------------------------------------------%%%
%%

Let us now describe the transformations of the polarization within
the cavity. We denote
%
%with $|V\rangle $ ($|H\rangle$) the linear polarization state
%orthogonal (parallel) to the plane of the cavity, and
%
with $\mathbb{Z}=|H\rangle \langle H|-|V\rangle \langle V|$ the
Pauli matrix with eigenstates $|V\rangle $ and $|H\rangle$. The
action of a cavity mirror on the polarization state is represented
by the unitary operator $\mathbb{M_Z}(\omega)=\exp[-{\rm
i}\phi_H(\omega)]|H\rangle \langle H|+ \exp[-{\rm
i}(\phi_V(\omega)+\pi)]|V\rangle \langle V|$~\cite{Azzam77} which,
apart from an unessential global phase factor, can be rewritten as
$\mathbb{M_Z}(\omega)=\mathbb{Z}\exp[-{\rm
i}\phi(\omega)\mathbb{Z}/2]$ and
$\phi(\omega)=\phi_H(\omega)-\phi_V(\omega)$ the relative phase
due to the polarization-reflectivity difference. The two plane
mirrors at 45$^\circ$ are characterized by the same
$\mathbb{M_Z}(\omega)$, while the third concave mirror of the
cavity is almost at normal incidence: for this mirror
$\phi(\omega)\simeq 0$ and therefore it acts as
$\mathbb{M_Z}\simeq \mathbb{Z}$. In order to compensate for this
operation we have inserted a waveplate $\mathbb{Z}$ in front of
the spherical mirror. The output polarization state after $n$
round-trips is given by the reduced density matrix obtained by
tracing over the frequency degree of freedom
\begin{eqnarray}\label{eq:outDMatrix}
    \hat{\rho}_{out}
        \!\!\!&=&\!\!\!\!\! \int d\mu_\omega\,
    \mathbb{U}[\phi(\omega)]^{n} |\pi \rangle_{in} \langle \pi|\mathbb{U}^{\dagger}[\phi(\omega)]^{n}
    \!\doteq\! \left [
        \begin{tabular}{cc}
            $\rho_{11}$ & $ \rho_{12}$\\
            $\rho_{12}^\ast$ & $\rho_{22}$\\
        \end{tabular}
    \right]\!\!,
\end{eqnarray}
where $\mathbb{U}[\phi(\omega)]$ is the unitary operator describing the polarization transformation after one cavity round-trip for a
given frequency component, and the $2 \times 2$ matrix is written in the $\{|H\rangle,|V\rangle\}$ basis.
The combined action of mirrors and compensating waveplate
$\mathbb{Z}$ yields
$\mathbb{U}[\phi(\omega)]^n=[\mathbb{M_Z}(\omega)\mathbb{M_Z}(\omega)]^n=\exp[-{\rm
i}n\phi(\omega)\mathbb{Z}]$.
The frequency average and the dispersive
properties of the 45$^\circ$-mirrors transform the pure input state $|\pi\rangle_{in}$ %\hat{\rho}_{in}$
into a mixed output state $\hat{\rho}_{out}$, with unmodified diagonal matrix elements but with
off-diagonal elements decaying to zero for an increasing number of cavity round-trips.

%%%---------------------------------------------------------%%%
%%%------------------ General Noise ---anticipated----------%%%
%%%---------------------------------------------------------%%%

This dephasing process is not the most general decoherence
affecting the polarization qubit. In the generic case decoherence
acts along an arbitrary, unknown, direction of the Bloch sphere,
rather than along the known $\hat{z}$-axis, affecting therefore
\textit{both} diagonal and off-diagonal elements of the density
matrix. We implemented the generic
%
%single-qubit
%
error model by placing in front of each plane mirror a
Soleil--Babinet (S-B) with axis at 45$^\circ$ with respect to the
cavity plane (see Fig.~\ref{fig:BB_Setup}). The action of the S-B
on the polarization state is described by $\mathbb{B_X}[\theta] =
\exp[-{\rm i}\theta \mathbb{X}/2]$, where $\mathbb{X}=|H\rangle
\langle V|+|V\rangle \langle H|$. The S-B together with a plane
mirror are described by the operator $\mathbb{N}(\omega,\theta) =
\mathbb{M_Z}(\omega)\mathbb{B_X}(\theta)$. The transformation of
the polarization state after the n$^{th}$-round trip in the
presence of the S--B is therefore given by
$\mathbb{U}[\phi(\omega),\theta]^n =
[\mathbb{N}(\omega,\theta)\,\mathbb{N}(\omega,\theta)]^{n}$.
The free evolution ($fe$) operator
$\mathbb{U}[\phi(\omega),\theta]$ can be rewritten as
$\exp\left[-i \alpha_{fe}(\omega,\theta)\,
\vec{s}_{fe}(\omega,\theta)\cdot \vec{\sigma}\right]$, where
$\vec{\sigma}=(\mathbb{X},\mathbb{Y},\mathbb{Z})$ is the vector of
the three Pauli matrices, and describes a rotation in the Bloch
sphere of an angle $2\,\alpha_{fe}(\omega,\theta)$ around the
direction individuated by $\vec{s}_{fe}(\omega,\theta)$. Therefore
by varying $\theta$ and the bandwidth of radiation spectrum, i.e.
the distribution of $\phi$, one implements the generic
polarization decoherence. $\alpha_{fe}(\omega,\theta)$ is given by
the implicit expression
\begin{equation}\label{eq:alphano}
\sin [\alpha_{fe}(\omega,\theta)/2]=\sin [\phi(\omega)/2]
\cos(\theta/2),
\end{equation}
and $\vec{s}_{fe}(\omega,\theta)=\{\sin\theta
[\cos\phi(\omega)-1],\sin\theta\sin\phi(\omega),(1+\cos\theta)
\sin\phi(\omega)\}$$/[2\sin\alpha_{fe}(\omega,\theta)]$.\\
%
%%%---------------------------------------------------------%%%
%%%--------------- FIRST EXPERIMENT ----NO BABINET----------%%%
%%%---------------------------------------------------------%%%
%
\begin{figure}[h!]
   \centering
   \includegraphics[width=.48\textwidth]{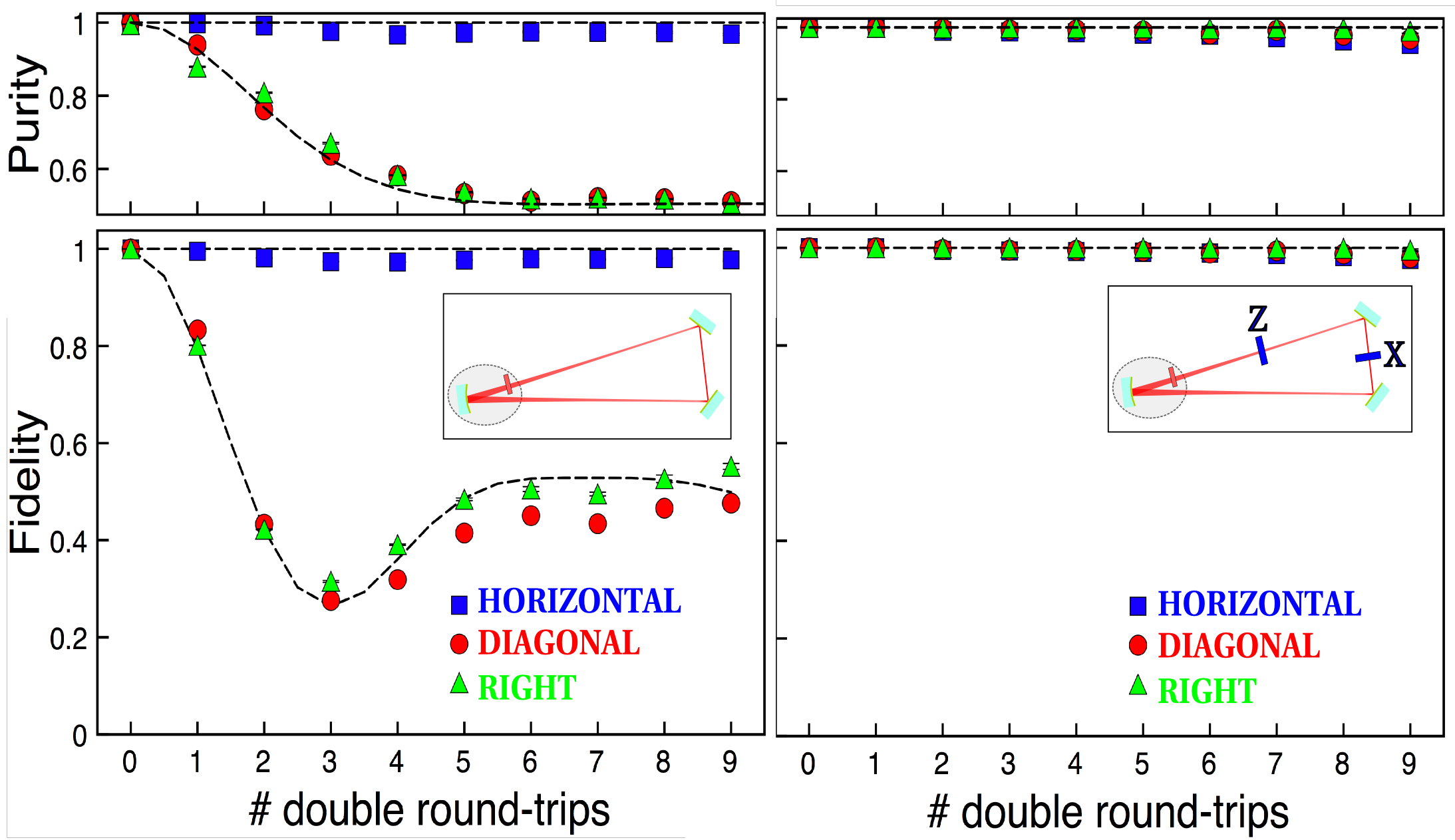}
   \caption{
      (Color online).
      Purity and fidelity versus the number of double round-trips
      for three different input polarization states (left without BB, right with BB) {\rm when decoherence is due only to the plane mirrors (no S--B
      included)}.
      Horizontal polarization $H$ is well preserved even without BB.
      On the contrary, if BB is not applied, the states $D$ and $R$
   quickly decay to the fully mixed state (purity and fidelity equal to $1/2$).
   {\rm The fitting curve has been obtained by considering a Gaussian spectrum for the pulse (see text)}.
   When BB is applied, both purity and fidelity remain very close
   to one for the whole duration of the photon storage in the cavity.
   Dashed lines fitting curves are obtained by numerical simulation with
   parameters $\sigma_\phi^{est} = 8.39\times10^{-2}\,{\rm rad}$ and $\phi_0 = 0.2182~{\rm rad}$.
   }
   \label{fig:BB_Results}
\end{figure}
\indent We have performed a first experiment without the S--B in
order to characterize the decohering properties of the cavity. The
results are illustrated in Fig.~\ref{fig:BB_Results}, where the
purity $\mathcal{P}={\rm Tr}(\rho_{{\rm out}}^2)$ and the fidelity
$\mathcal{F}=\,_{in}\langle \pi|\rho_{{\rm out}}|\pi\rangle_{in}$
of the output polarization state are plotted versus the number of
BB cycles (double round-trips), for the input polarization states
$H$, $D$ ($45^{\circ}$ linear polarization) and $R$ (right circular polarization state). 
The density matrices are
evaluated by maximum-likelihood estimation from the histogram
obtained with the TAC/MCA system for different settings of the
waveplates in the tomography apparatus. Ten time-bins around each
peak of the histograms, corresponding to an integration time
window of 1~ns, have been summed for evaluating the detector
counts for the corresponding round-trip. When BB is not performed,
polarizations $D$ and $R$ decay to the fully unpolarized state
($\mathcal{P}=\mathcal{F}=1/2$), while $H$, being an eigenstate of
$\mathbb{Z}$, is unaffected by decoherence. On the contrary,
polarization decoherence is completely suppressed when BB is
applied. The BB is realized by adding two control operations
within the cavity: a second ${\mathbb Z}$ waveplate before the
spherical mirror, and a S--B with axis at 45$^\circ$ with respect
to the cavity plane and delay equal to $\lambda/2$ in the short
arm of the cavity, acting therefore as $\mathbb{X}$. The two
controls implement every two cavity round-trips the full
Pauli-group decoupling of a qubit~\cite{Viola99,Massar2007}. The
transformation after the n$^{th}$-round trip then becomes
$\mathbb{U}[\phi(\omega)]^{n}
        = [\mathbb{Z}\,\mathbb{M_{Z}}(\omega)\,
        \mathbb{X}\,\mathbb{M_{Z}}(\omega)]^{n}
        =({\rm i}\mathbb{Y})^{n}\,;$
therefore, for even $n$, the polarization transformation is
proportional to the identity operator, thus implying a
\textit{perfect preservation of every input polarization state},
i.e., a complete suppression of decoherence \cite{notes}. This is
well verified for $H$, $D$ and $R$ polarizations in
Fig.~\ref{fig:BB_Results}.\\
\indent The decay of the purity for $D$ and $R$ input states
allows us to get a quantitative estimate of the polarization
decoherence caused by the two plane mirrors. The output
polarization state is given by the frequency average of
Eq.~(\ref{eq:outDMatrix}) and can be written as
$\rho_{out}=[I+\vec{P}_{out}\cdot \vec{\sigma}]/2$, where
$\vec{P}_{out}$ is the corresponding Bloch vector, so that
$\mathcal{P}=[1+|\vec{P}_{out}|^2]/2$. The frequency average can
be treated by assuming $\left|\mathcal{E}(\omega)\right|^2 =
(\pi\sigma^2_\omega)^{-1/2}\, \exp[-(\omega -
\omega_0)^2/\sigma_\omega^2]$, where $\sigma_\omega$ represents
the bandwidth of the radiation spectrum centered in $\omega_0$.
The frequency dependence enters through the relative phase
$\phi(\omega)$ due to the polarization-reflectivity difference at
the plane mirrors. This dependence is well described by the linear
relation $\phi (\omega) \simeq \phi_0+\tau (\omega - \omega_0)$,
where $\phi_0=\phi(\omega_0)$, so that the integral of
Eq.~(\ref{eq:outDMatrix}) becomes an average over a Gaussian
measure with standard deviation $\sigma_\phi = \tau
\sigma_\omega$. The output purity without BB after $n$ cavity
round-trips for either $D$ and $R$ input polarizations is then
given by $\mathcal{P}=[1 +\exp(-2n^2\sigma^2_\phi)]/2$, which can
be used for a best fit on the experimental data, giving
$\sigma_\phi^{est} = (8.39\pm 0.03)\times 10^{-2}\,{\rm rad}$.
%

%%%---------------------------------------------------------%%%
%%%------------------ Second experiment with Babinets -------%%%
%%%---------------------------------------------------------%%%

We have performed a second experiment for the more general model
of decoherence
%
%in the case when polarization decoherence acts along an unknown
%direction of the Bloch sphere,
%
by inserting along the optical path a S--B in front of each plane
mirror. By changing the delay of the S--B, one changes $\theta$
and therefore the orientation of the decoherence axis. Pauli-group
decoupling is again realized every two round-trips by adding the
BB operations $\mathbb{X}$ and $\mathbb{Z}$ in the
cavity~\cite{Viola99,Massar2007}. The overall transformation after
the n$^{th}$-round trip for a given frequency component is now
given by
$\mathbb{U}[\phi(\omega),\theta]^n =[\mathbb{Z}\,\mathbb{N}(\omega,\theta)\,
        \mathbb{X}\,\mathbb{N}(\omega,\theta)]^{n}\,,$
which can be rewritten, as in the free evolution case, as a rotation in
the Bloch sphere, $\exp[-i\alpha_{bb}(\omega,\theta)\,\vec{s}_{bb}(\omega,\theta)\cdot
\vec{\sigma}]$, with
\begin{equation}\label{eq:implicitalph}
\cos \alpha_{bb}(\omega,\theta)=-(\sin\phi(\omega)\,\sin\theta)/2,
\end{equation}
\begin{figure}[h!]
   \centering
   \includegraphics[width=.48\textwidth]{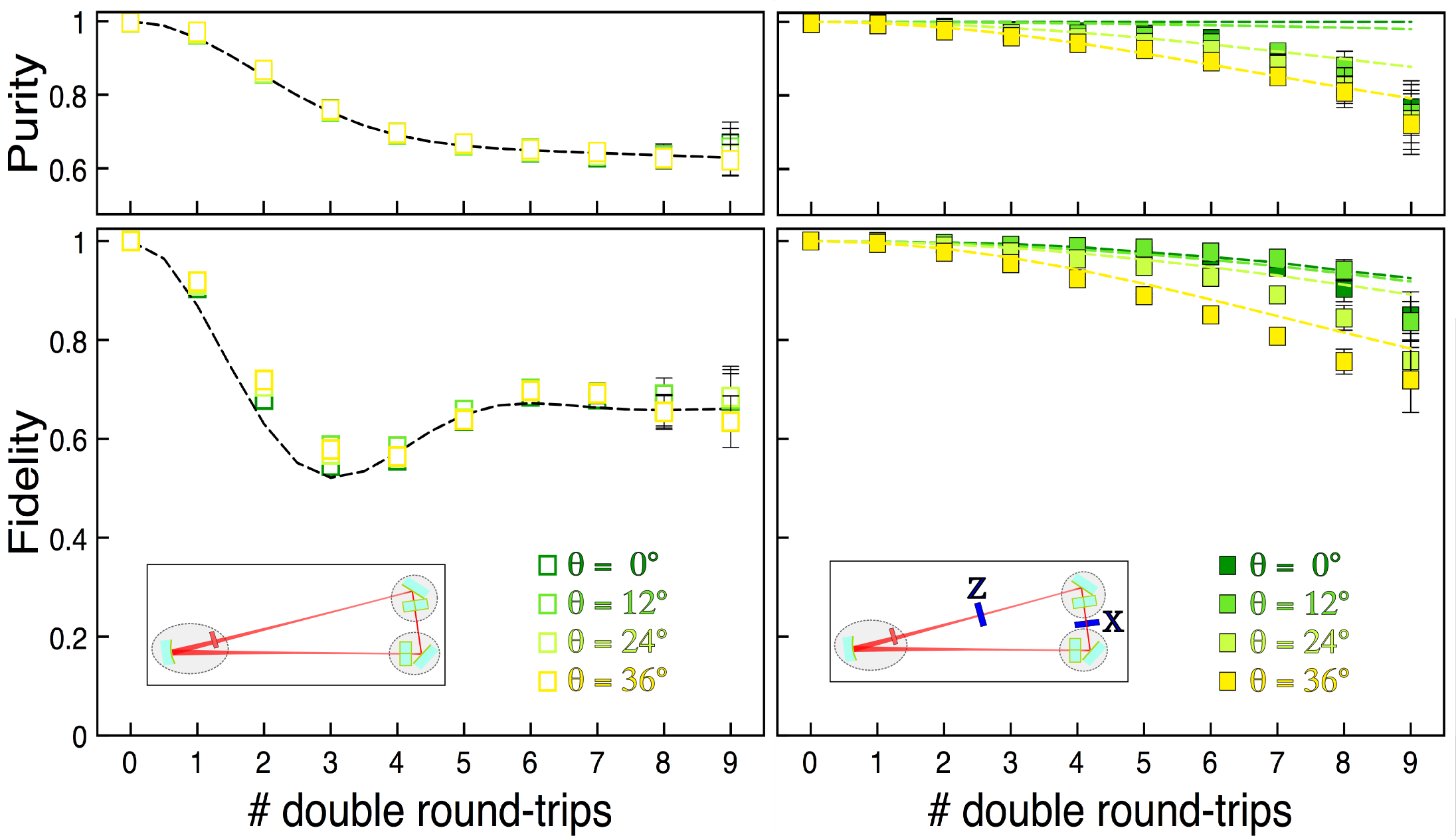}
   \caption{
    (Color online).
Purity and fidelity, averaged over the whole Bloch sphere, for
generic decoherence, for different phase angles, $\theta$, of the
S--B (left without BB, right with BB). For each orientation of the
decoherence-axis the output purity and fidelity with BB are
significantly higher than the corresponding value without BB.
Dashed lines fitting curves obtained by numerical simulation with
$\sigma_\phi^{est} = 8.39\times10^{-2}\,{\rm rad}$ and $\phi_0 = 0.2182~{\rm rad}$,
as in Figure~\ref{fig:BB_Results}.
   }
   \label{fig:BB_GenNoise}
\end{figure}
and $\vec{s}_{bb}(\omega,\theta)=$ $\{-\sin\phi(\omega)$
$\sin^2(\theta/2)$, $1-2\sin^2(\theta/2)$
$\sin^2[\phi(\omega)/2]$, $-\sin\theta\sin^2[\phi(\omega)/2]\}/$
$[\sin\alpha_{bb}(\omega,\theta)]$. Figure~\ref{fig:BB_GenNoise}
shows the evolution of the purity and the fidelity, averaged over
the whole Bloch sphere, under this general decoherence model, both
with and without BB decoupling.
Each curve corresponds to a different orientation $\theta$ of the
S-B. BB again inhibits decoherence because, for each orientation
of the decoherence-axis, both the average purity and the average
fidelity in the presence of BB are significantly higher than the
corresponding value without BB. The curve for $\theta =0$
reproduces the almost perfect preservation of the previous
experiment, but polarization protection worsens for increasing
$\theta$. This can be understood by performing the average of
Eq.~(\ref{eq:outDMatrix}) under the assumption that the pulse is
not too broad, $\sigma_{\phi} \ll \pi$, which is well verified in
our experiment. In this case, $\alpha_{j}$ and $\vec{s}_{j}$
($j=fe,bb$) do not vary appreciably over the range of relevant
phase shifts $\phi$ and one can approximate $\vec{s}_j(\phi)$ with
its value at the pulse center $\vec{s}_j(\phi_0)$, while
$\alpha_j(\phi)$ can be approximated by its first-order expansion
around $\phi_0$, $\alpha_j(\phi)\simeq \alpha_j^0+\dot{\alpha}_j^0
(\phi-\phi_0)$. From the output density matrix one then derives
the expression of $\mathcal{P}$ and $\mathcal{F}$ as a function of
$n$. In the limit $n \to \infty$ both $\mathcal{P}$ and
$\mathcal{F}$ tends to $\{1+[\vec{P}_{in}\cdot
\vec{s}(\phi_0)]^2\}/2$, which corresponds to the existence of a
``pointer'' basis unaffected by decoherence \cite{Zurek2003}
(analogous to the principal states of polarization in fibers
\cite{Poole88}), formed by the two states with Bloch vector equal
to $\pm \vec{s}(\phi_0)$. As a consequence, when averaged over the
initial state, purity and fidelity tends to $2/3$. However, the
interesting regime of our experiment is the one corresponding to a
small round-trip number $n$. In that regime our ring-cavity well
mimics a portion of a quantum communication channel with constant
dispersive properties. For small $n$ one has 
$1-\mathcal{F}\simeq (1-\mathcal{P})/2\simeq n^2
\left[\dot{\alpha}_j^0 \sigma_{\phi}\right]^2 \{
1-[\vec{P}_{in}\cdot \vec{s}_j(\phi_0)]^2\}/2$,
with $j=\{fe,bb\}$, showing that the smaller $\dot{\alpha}^0$ the
better the decoherence suppression. BB Pauli-group decoupling acts
just by decreasing $ |\dot{\alpha}^0 |$: by using
Eqs.~(\ref{eq:alphano})-(\ref{eq:implicitalph}) and the fact that
$\phi_0$ is quite small in our experiment, one gets
$\dot{\alpha}_{fe}^0 \simeq \cos(\theta/2)$,
$\dot{\alpha}_{bb}^0 \simeq \sin(\theta/2)\cos(\theta/2)$,
showing that it is always $\dot{\alpha}_{fe}^0 >
\dot{\alpha}_{bb}^0$ and therefore that BB better preserves the
polarization qubit for any orientation of the decoherence axis,
confirming its applicability to otpical fibers polarization control.
These expressions also explain the perfect preservation of the polarization qubit of Fig.~\ref{fig:BB_Results}: the latter refers to $\theta
=0$, implying $\dot{\alpha}_{fe}^0 =1$ and $\dot{\alpha}_{bb}^0=0$.
\\\indent
In conclusion we have provided a strong experimental
evidence of the usefulness of BB Pauli-group decoupling in
protecting a flying qubit against generic polarization
decoherence. Our results suggest that BB may prove helpful to increase the fidelity of a one-way,
polarization-based, quantum transmission as well as the coherence
of a polarization qubit, over distances much longer than currently
achieved. In this respect one remark is in order. The
BB decoupling technique adopted by us
is fundamentally different from the technique of Passive
Compensation (PC), often used in quantum communication
schemes~\cite{Gisin2007}. PC exploits the retracing property of a
beam traveling back and forth between two mirrors to revert on
every ``backward path'' the dephasing introduced by dispersive
elements during each ``forward path''. Therefore, PC only works
for \textit{two-way} channels, as exemplified by the
Faraday-PC~\cite{Martinelli1989}. Also the demonstration
in~\cite{Berglund00} works only for a linear cavity and requires
PC as an essential tool. On the contrary, in the ring-cavity of
our setup, the photon always travels in the forward direction,
thus proving the BB effectiveness in the more demanding task of
flying qubit coherence-maintenance on a \textit{one-way} channel.
\\\indent
This work has been partly supported by the EC IP QAP, Contract No. 015848. We thank L. Viola and G. Lo Bianco for illuminating discussions.
%

%\bibliography{PRL_20082212}
%\bibliography{/Users/gianx/MyStuff/LaTexStuff/MyBib/MyBib}
%\bibliography{MyBib}

%\bibliography{PRL_20090620.bbl}
%\bibliography{/media/sda5/Gianx/Camerino/Papers/MyBib/MyBib}

\begin{thebibliography}{30}

\expandafter\ifx\csname
natexlab\endcsname\relax\def\natexlab#1{#1}\fi
\expandafter\ifx\csname bibnamefont\endcsname\relax
  \def\bibnamefont#1{#1}\fi
\expandafter\ifx\csname bibfnamefont\endcsname\relax
  \def\bibfnamefont#1{#1}\fi
\expandafter\ifx\csname citenamefont\endcsname\relax
  \def\citenamefont#1{#1}\fi
\expandafter\ifx\csname url\endcsname\relax
  \def\url#1{\texttt{#1}}\fi
\expandafter\ifx\csname
urlprefix\endcsname\relax\def\urlprefix{URL }\fi
\providecommand{\bibinfo}[2]{#2}
\providecommand{\eprint}[2][]{\url{#2}}

\bibitem[{\citenamefont{Ernst et~al.}(1987)\citenamefont{Ernst, Bodenhausen,
  and Wokaun}}]{Ernst1987}
\bibinfo{author}{\bibfnamefont{R.}~\bibnamefont{Ernst}},
  \bibinfo{author}{\bibfnamefont{G.}~\bibnamefont{Bodenhausen}},
  \bibnamefont{and} \bibinfo{author}{\bibfnamefont{A.}~\bibnamefont{Wokaun}},
  \textit{\bibinfo{title}{Principles of Nuclear Magnetic Resonance in One and Two
  Dimensions}} (\bibinfo{publisher}{Clarendon Press, Oxford},
  \bibinfo{year}{1987}).

\bibitem[{\citenamefont{Shor}(1995)}]{Shor1995}
\bibinfo{author}{\bibfnamefont{P.~W.} \bibnamefont{Shor}},
  \bibinfo{journal}{Phys. Rev. A} \textbf{\bibinfo{volume}{52}},
  \bibinfo{pages}{R2493} (\bibinfo{year}{1995}).

\bibitem[{\citenamefont{Steane}(1999)}]{Steane1999}
\bibinfo{author}{\bibfnamefont{A.~M.} \bibnamefont{Steane}},
  \bibinfo{journal}{Nature} \textbf{\bibinfo{volume}{399}},
  \bibinfo{pages}{124} (\bibinfo{year}{1999}).

\bibitem[{\citenamefont{Chiaverini et~al.}(2004)\citenamefont{Chiaverini,
  Leibfried, Schaetz, Barrett, Blakestad, Britton, Itano, Jost, Knill, Langer
  et~al.}}]{Chiaverini2004}
\bibinfo{author}{\bibfnamefont{J.}~\bibnamefont{Chiaverini}},
  \bibnamefont{et~al.}, \bibinfo{journal}{Nature}
  \textbf{\bibinfo{volume}{432}}, \bibinfo{pages}{602} (\bibinfo{year}{2004}).

\bibitem[{\citenamefont{Boulant et~al.}(2005)\citenamefont{Boulant, Viola,
  Fortunato, and Cory}}]{Boulant2005}
\bibinfo{author}{\bibfnamefont{N.}~\bibnamefont{Boulant}},
  \bibnamefont{et~al.},
  \bibinfo{journal}{Phys. Rev. Lett.} \textbf{\bibinfo{volume}{94}},
  \bibinfo{eid}{130501} (\bibinfo{year}{2005}).

\bibitem[{\citenamefont{Zanardi and Rasetti}(1997)}]{Zanardi1997}
\bibinfo{author}{\bibfnamefont{P.}~\bibnamefont{Zanardi}} \bibnamefont{and}
  \bibinfo{author}{\bibfnamefont{M.}~\bibnamefont{Rasetti}},
  \bibinfo{journal}{Phys. Rev. Lett.} \textbf{\bibinfo{volume}{79}},
  \bibinfo{pages}{3306} (\bibinfo{year}{1997}).

\bibitem[{\citenamefont{Kielpinski et~al.}(2001)\citenamefont{Kielpinski,
  Meyer, Rowe, Sackett, Itano, Monroe, and Wineland}}]{Kielpinski2001}
\bibinfo{author}{\bibfnamefont{D.}~\bibnamefont{Kielpinski}},
  \bibnamefont{et~al.},
  \bibinfo{journal}{Science} \textbf{\bibinfo{volume}{291}},
  \bibinfo{pages}{1013} (\bibinfo{year}{2001}).

\bibitem[{\citenamefont{Prevedel et~al.}(2007)\citenamefont{Prevedel, Tame,
  Stefanov, Paternostro, Kim, and Zeilinger}}]{Prevedel2007}
\bibinfo{author}{\bibfnamefont{R.}~\bibnamefont{Prevedel}},
  \bibnamefont{et~al.},
  \bibinfo{journal}{Phys. Rev. Lett.} \textbf{\bibinfo{volume}{99}},
  \bibinfo{eid}{250503} (\bibinfo{year}{2007}).

\bibitem[{\citenamefont{Vitali et~al.}(1997)\citenamefont{Vitali, Tombesi, and
  Milburn}}]{Vitali1997}
\bibinfo{author}{\bibfnamefont{D.}~\bibnamefont{Vitali}},
  \bibinfo{author}{\bibfnamefont{P.}~\bibnamefont{Tombesi}}, \bibnamefont{and}
  \bibinfo{author}{\bibfnamefont{G.~J.} \bibnamefont{Milburn}},
  \bibinfo{journal}{Phys. Rev. Lett.} \textbf{\bibinfo{volume}{79}},
  \bibinfo{pages}{2442} (\bibinfo{year}{1997}).

\bibitem[{\citenamefont{Viola and Lloyd}(1998)}]{Viola1998}
\bibinfo{author}{\bibfnamefont{L.}~\bibnamefont{Viola}} \bibnamefont{and}
  \bibinfo{author}{\bibfnamefont{S.}~\bibnamefont{Lloyd}},
  \bibinfo{journal}{Phys. Rev. A} \textbf{\bibinfo{volume}{58}},
  \bibinfo{pages}{2733} (\bibinfo{year}{1998}).

\bibitem[{\citenamefont{Viola et~al.}(1999)\citenamefont{Viola, Knill, and
  Lloyd}}]{Viola99}
\bibinfo{author}{\bibfnamefont{L.}~\bibnamefont{Viola}},
  \bibinfo{author}{\bibfnamefont{E.}~\bibnamefont{Knill}}, \bibnamefont{and}
  \bibinfo{author}{\bibfnamefont{S.}~\bibnamefont{Lloyd}},
  \bibinfo{journal}{Phys. Rev. Lett.} \textbf{\bibinfo{volume}{82}},
  \bibinfo{pages}{2417} (\bibinfo{year}{1999}).

\bibitem[{\citenamefont{Vitali and Tombesi}(1999)}]{Vitali1999}
\bibinfo{author}{\bibfnamefont{D.}~\bibnamefont{Vitali}} \bibnamefont{and}
  \bibinfo{author}{\bibfnamefont{P.}~\bibnamefont{Tombesi}},
  \bibinfo{journal}{Phys. Rev. A} \textbf{\bibinfo{volume}{59}},
  \bibinfo{pages}{4178} (\bibinfo{year}{1999}).

\bibitem[{\citenamefont{Zanardi}(1999)}]{Zanardi1999}
\bibinfo{author}{\bibfnamefont{P.}~\bibnamefont{Zanardi}},
  \bibinfo{journal}{Phys. Lett. A} \textbf{\bibinfo{volume}{258}},
  \bibinfo{pages}{77} (\bibinfo{year}{1999}).

\bibitem[{\citenamefont{Viola et~al.}(2001)\citenamefont{Viola, Fortunato,
  Pravia, Knill, Laflamme, and Cory}}]{Viola2001}
\bibinfo{author}{\bibfnamefont{L.}~\bibnamefont{Viola}},
  \bibnamefont{et~al.},
  \bibinfo{journal}{Science} \textbf{\bibinfo{volume}{293}},
  \bibinfo{pages}{2059} (\bibinfo{year}{2001}).

\bibitem[{\citenamefont{Kofman and Kurizki}(2004)}]{Kofman2004}
\bibinfo{author}{\bibfnamefont{A.~G.} \bibnamefont{Kofman}} \bibnamefont{and}
  \bibinfo{author}{\bibfnamefont{G.}~\bibnamefont{Kurizki}},
  \bibinfo{journal}{Phys. Rev. Lett.} \textbf{\bibinfo{volume}{93}},
  \bibinfo{pages}{130406} (\bibinfo{year}{2004}).

\bibitem[{\citenamefont{Fraval et~al.}(2005)\citenamefont{Fraval, Sellars, and
  Longdell}}]{Fraval2005}
\bibinfo{author}{\bibfnamefont{E.}~\bibnamefont{Fraval}},
  \bibinfo{author}{\bibfnamefont{M.~J.} \bibnamefont{Sellars}},
  \bibnamefont{and} \bibinfo{author}{\bibfnamefont{J.~J.}
  \bibnamefont{Longdell}}, \bibinfo{journal}{Phys. Rev. Lett.}
  \textbf{\bibinfo{volume}{95}}, \bibinfo{pages}{030506}
  (\bibinfo{year}{2005}).

\bibitem[{\citenamefont{Facchi et~al.}(2005)\citenamefont{Facchi, Tasaki,
  Pascazio, Nakazato, Tokuse, and Lidar}}]{Facchi2005}
\bibinfo{author}{\bibfnamefont{P.}~\bibnamefont{Facchi}},
  \bibnamefont{et~al.},
  \bibinfo{journal}{Phys. Rev. A}
  \textbf{\bibinfo{volume}{71}}, \bibinfo{eid}{022302}
  (\bibinfo{year}{2005}).

\bibitem[{\citenamefont{Morton et~al.}(2006)\citenamefont{Morton, Tyryshkin,
  Ardavan, Benjamin, Porfyrakis, Lyon, and Briggs}}]{Morton2006}
\bibinfo{author}{\bibfnamefont{J.~J.~L.} \bibnamefont{Morton}},
  \bibnamefont{et~al.},
  \bibinfo{journal}{Nature Physics} \textbf{\bibinfo{volume}{2}},
  \bibinfo{pages}{40} (\bibinfo{year}{2006}).

\bibitem[{\citenamefont{Agarwal et~al.}(2001)\citenamefont{Agarwal, Scully, and
  Walther}}]{Agarwal2001a}
\bibinfo{author}{\bibfnamefont{G.~S.} \bibnamefont{Agarwal}},
  \bibinfo{author}{\bibfnamefont{M.~O.} \bibnamefont{Scully}},
  \bibnamefont{and} \bibinfo{author}{\bibfnamefont{H.}~\bibnamefont{Walther}},
  \bibinfo{journal}{Phys. Rev. Lett.} \textbf{\bibinfo{volume}{86}},
  \bibinfo{pages}{4271} (\bibinfo{year}{2001}).

\bibitem[{\citenamefont{Search and Berman}(2000)}]{Search2000a}
\bibinfo{author}{\bibfnamefont{C.}~\bibnamefont{Search}} \bibnamefont{and}
  \bibinfo{author}{\bibfnamefont{P.~R.} \bibnamefont{Berman}},
  \bibinfo{journal}{Phys. Rev. Lett.} \textbf{\bibinfo{volume}{85}},
  \bibinfo{pages}{2272} (\bibinfo{year}{2000}).

\bibitem[{\citenamefont{Vitali and Tombesi}(2001)}]{Vitali2002}
\bibinfo{author}{\bibfnamefont{D.}~\bibnamefont{Vitali}} \bibnamefont{and}
  \bibinfo{author}{\bibfnamefont{P.}~\bibnamefont{Tombesi}},
  \bibinfo{journal}{Phys. Rev. A} \textbf{\bibinfo{volume}{65}},
  \bibinfo{pages}{012305} (\bibinfo{year}{2001}).

\bibitem[{\citenamefont{A.~Galtarossa and Tambosso}(2000)}]{Galtarossa2000}
\bibinfo{author}{\bibfnamefont{M.~S.} \bibnamefont{A.~Galtarossa},
  \bibfnamefont{L.~Palmieri}} \bibnamefont{and}
  \bibinfo{author}{\bibfnamefont{T.}~\bibnamefont{Tambosso}},
  \bibinfo{journal}{Opt. Lett.} \textbf{\bibinfo{volume}{25}},
  \bibinfo{pages}{384} (\bibinfo{year}{2000}).

\bibitem[{\citenamefont{Massar and Popescu}(2007)}]{Massar2007}
\bibinfo{author}{See}
\bibinfo{author}{\bibfnamefont{S.}~\bibnamefont{Massar}} \bibnamefont{and}
  \bibinfo{author}{\bibfnamefont{S.}~\bibnamefont{Popescu}},
  \bibinfo{journal}{New J. Phys.} \textbf{\bibinfo{volume}{9}},
  \bibinfo{pages}{158} (\bibinfo{year}{2007}), \bibinfo{author}{where they also
  discuss a model for spun fibers.}

%discuss the difference with spinning the fiber}.

\bibitem[{\citenamefont{James et~al.}(2001)\citenamefont{James, Kwiat, Munro,
  and White}}]{James2001}
\bibinfo{author}{\bibfnamefont{D.~F.~V.} \bibnamefont{James}},
  \bibnamefont{et~al.},
  \bibinfo{journal}{Phys. Rev. A} \textbf{\bibinfo{volume}{64}},
  \bibinfo{pages}{052312} (\bibinfo{year}{2001}).

\bibitem[{\citenamefont{Azzam and Bashara}(1977)}]{Azzam77}
\bibinfo{author}{\bibfnamefont{R.~M.~A.} \bibnamefont{Azzam}} \bibnamefont{and}
  \bibinfo{author}{\bibfnamefont{N.~M.} \bibnamefont{Bashara}},
  \textit{\bibinfo{title}{Ellipsometry and Polarized Light}}
  (\bibinfo{publisher}{North-Holland}, \bibinfo{address}{Amsterdam},
  \bibinfo{year}{1977}).

\bibitem[{\citenamefont{}()}]{notes}
\bibinfo{author}{ In this first experiment the decoherence axis is
known to be along $\hat{z}$ and Pauli-group decoupling coincides
with Carr-Purcell decoupling \cite{Viola1998} involving only the
$\mathbb{X}$ control.}

\bibitem[{\citenamefont{Zurek}(2003)}]{Zurek2003}
\bibinfo{author}{\bibfnamefont{W.~H.} \bibnamefont{Zurek}},
  \bibinfo{journal}{Rev. Mod. Phys.} \textbf{\bibinfo{volume}{75}},
  \bibinfo{pages}{715} (\bibinfo{year}{2003}).

\bibitem[{\citenamefont{Poole}(1988)}]{Poole88}
\bibinfo{author}{\bibfnamefont{C.~D.} \bibnamefont{Poole}},
  \bibinfo{journal}{Opt. Lett.} \textbf{\bibinfo{volume}{13}},
  \bibinfo{pages}{687} (\bibinfo{year}{1988}).

\bibitem[{\citenamefont{Gisin and Thew}(2007)}]{Gisin2007}
\bibinfo{author}{\bibfnamefont{N.}~\bibnamefont{Gisin}} \bibnamefont{and}
  \bibinfo{author}{\bibfnamefont{R.}~\bibnamefont{Thew}}, \bibinfo{journal}{Nature
  Photonics} \textbf{\bibinfo{volume}{1}}, \bibinfo{pages}{165}
  (\bibinfo{year}{2007}).

\bibitem[{\citenamefont{Martinelli}(1989)}]{Martinelli1989}
\bibinfo{author}{\bibfnamefont{M.}~\bibnamefont{Martinelli}},
  \bibinfo{journal}{Opt. Commun.} \textbf{\bibinfo{volume}{72}},
  \bibinfo{pages}{341} (\bibinfo{year}{1989}).

\bibitem[{\citenamefont{Berglund}(2000)}]{Berglund00}
\bibinfo{author}{\bibfnamefont{A.}~\bibnamefont{Berglund}},
  \bibinfo{journal}{quant-ph/0010001}  (\bibinfo{year}{2000}).

\end{thebibliography}

\end{document}